\documentclass[12pt]{article}

\usepackage{cite,ulem,epsfig,amsmath,booktabs}

\newcommand{\mycaption}[1]{\caption{\sl #1}}

\newcommand{\anc}{\rule{0mm}{0mm}}
\newcommand{\Eslash}{{\not{\!\!E}}}

\newcommand{\LL}{{\rm L}}
\newcommand{\st}{{\tilde{\tau}}}

\newcommand{\sq}{{\tilde{q}}}
\newcommand{\sqR}{{\tilde{q}_{\rm R}}}
\newcommand{\sqL}{{\tilde{q}_{\rm L}}}

\newcommand{\suR}{{\tilde{u}_{\rm R}}}
\newcommand{\suL}{{\tilde{u}_{\rm L}}}

\newcommand{\sdR}{{\tilde{d}_{\rm R}}}
\newcommand{\sdL}{{\tilde{d}_{\rm L}}}
\newcommand{\go}{\tilde{g}}
\newcommand{\cha}{\tilde{\chi}}
\newcommand{\neu}{\tilde{\chi}^0}
\newcommand{\mcha}[1]{m_{\tilde{\chi}^\pm_{#1}}}
\newcommand{\mneu}[1]{m_{\tilde{\chi}^0_{#1}}}

\newcommand{\gev}{{\rm \ GeV}}

\begin{document}
\thispagestyle{empty}

\def\thefootnote{\fnsymbol{footnote}}

\begin{flushright}
DESY 07-035                 \\
FERMILAB--Pub--07/035--T    \\
PSI--PR--07--01             \\
ZU--TH 07/07
\end{flushright}

\vspace{1cm}

\begin{center}

{\Large\sc {\bf Examining the Identity of Yukawa with Gauge \\[3mm]
                Couplings in Supersymmetric QCD at LHC}}
\\[3.5em]
{\large\sc
A.~Freitas$^{1}$,
P.~Skands$^{2}$,
M.~Spira$^{3}$, and 
P.~M.~Zerwas$^{4}$
}

\vspace*{1cm}
 
{\sl
$^1$ Institut f\"ur Theoretische Physik,
        Universit\"at Z\"urich, \\ Winterthurerstrasse 190, CH-8057
        Z\"urich, Switzerland

\vspace*{0.4cm}

$^2$ Theoretical Physics, Fermi National Accelerator Laboratory,\\ P.\
O.\ Box 500, Batavia, IL-60510, USA

\vspace*{0.4cm}

$^3$ Paul Scherrer Institut, CH-5232 Villigen PSI, Switzerland

\vspace*{0.4cm}

$^4$ Deutsches Elektronensynchrotron DESY, D-22603 Hamburg, Germany

\vspace*{0.4cm}

}

\end{center}


\vspace*{1cm}

\begin{abstract}
The identity of the quark-squark-gluino Yukawa coupling
with the corresponding quark-quark-gluon QCD coupling in 
supersymmetric theories can be examined experimentally 
at the Large Hadron Collider (LHC). Extending earlier investigations of
like-sign di-lepton final states, we include jets in the analysis
of the minimal supersymmetric standard model, adding
squark-gluino and gluino-pair production to 
squark-pair production. Moreover
we expand the method towards model-independent analyses 
which cover  more general scenarios. In all cases, squark decays 
to light charginos and neutralinos persist to play a 
dominant role.  
\end{abstract}

\def\thefootnote{\arabic{footnote}}
\setcounter{page}{0}
\setcounter{footnote}{0}

\newpage


\section{Introduction}

A characteristic consequence of supersymmetry is the 
identity of the quark-squark-gluino Yukawa coupling with the
quark-quark-gluon coupling in SUSY-QCD \cite{WessZumino}:
\begin{equation}
\hat{g}_{\rm s}(q \tilde{q} \tilde{g}) = {g}_{\rm s}(qqg)
\end{equation}
which persists to hold after the supersymmetry is broken softly.
Similar identities are predicted in other sectors 
of the supersymmetric extension of the Standard Model. 

Pair production of sleptons and charginos/neutralinos in $e^+e^-$
collisions has been investigated to test the identity 
of Yukawa with gauge couplings in the non-coloured electroweak 
sector, {\it cf.} Refs.$\,$\cite{ewyuk,CKMZ,ewyukc}. The equality of 
electroweak Yukawa and gauge couplings is expected to be studied 
at the per-cent to per-mil level by these experiments. 
In a recent report, Ref.~\cite{quigleys}, the measurement of like-sign 
di-leptons has been examined to study the squark-pair production 
processes $q q \to \tilde{q} \tilde{q}$. Since these processes are
mediated solely by the exchange of gluinos, they provide, in principle, 
the most appealing measurement of the Yukawa coupling in supersymmetric 
QCD. All branching ratios in the decay cascade 
$\tilde{q} \to \tilde{\chi}_1^\pm \to \tilde{\ell}$ have been assumed to be 
known in evaluating the $\tilde{q} \tilde{q}$ pair cross section which
involves the Yukawa coupling. 

In this note we extend the analysis in two directions. First we will include
the measurement of jets in like-sign di-lepton events, motivated by the 
large production cross sections for gluinos \cite{Hop} which
significantly enlarge the like-sign di-lepton ensemble
through instantaneous decays to real or virtual squarks 
[see also Ref.~\cite{noj07}]. In a second step 
we will investigate experimental scenarios in which only the properties of 
light gaugino-type charginos and neutralinos have been pre-determined
either at LHC or at an $e^+e^-$ linear collider \cite{LC}. This
method, within a wide variety of theories, does not rely, in particular,
on the measurement 
of the decay branching ratios of heavy squarks. Instead of studying the 
absolute size of the cross sections, we will analyse ratios 
of like-sign di-lepton cross sections and the evolution 
of these cross sections with the transverse momenta 
of additional jets. Including jets with varying transverse momenta
in the analysis changes the relative weight 
of the subprocesses with different dependence
on the Yukawa coupling. Therefore, relating the cross sections for
different final state configurations, 
the overall normalisation can be eliminated from the analysis
and the Yukawa coupling can be determined without {\it a priori} knowledge 
of the branching ratios for $\tilde{q}$ decays.

\section{Squark and gluino production at the LHC}

\begin{figure}
\begin{tabular}{@{}rlr@{}}
\psfig{figure=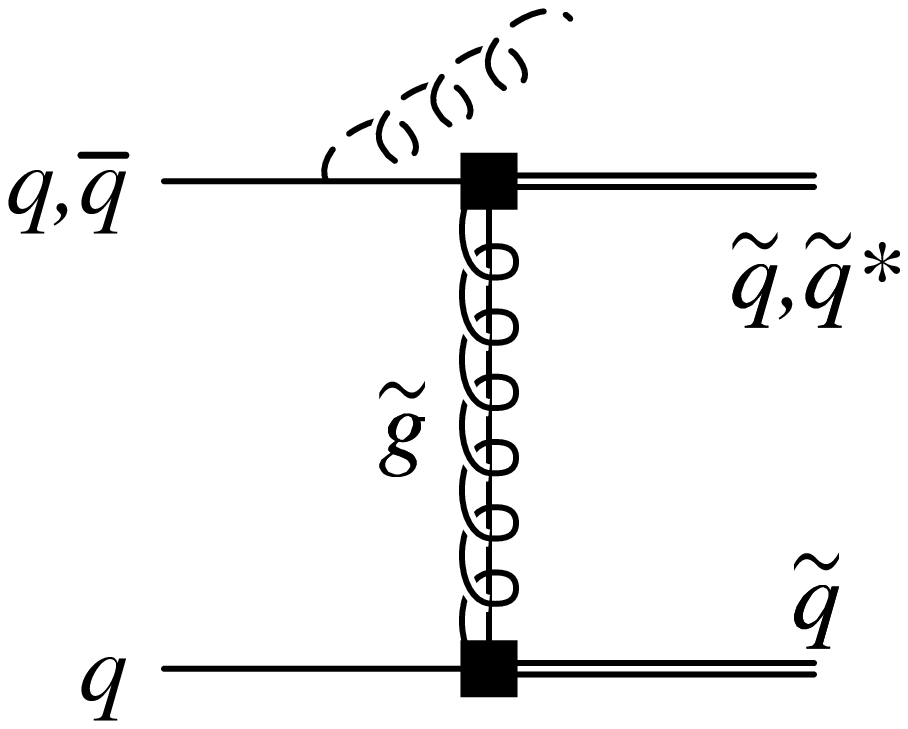, height=3.5cm} \anc &
\psfig{figure=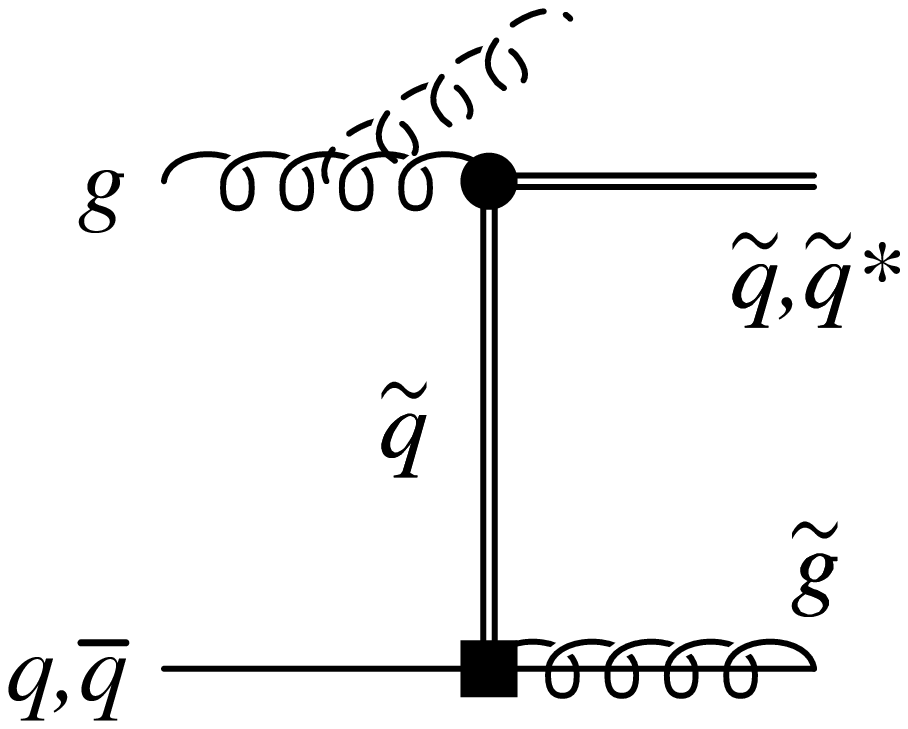, height=3.5cm}&
\psfig{figure=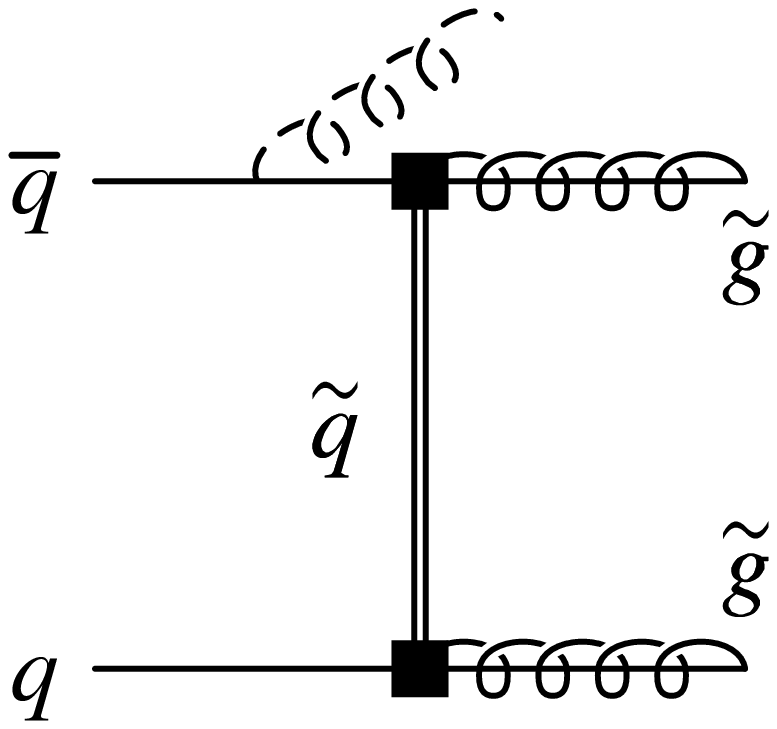, height=3.5cm}\\
\psfig{figure=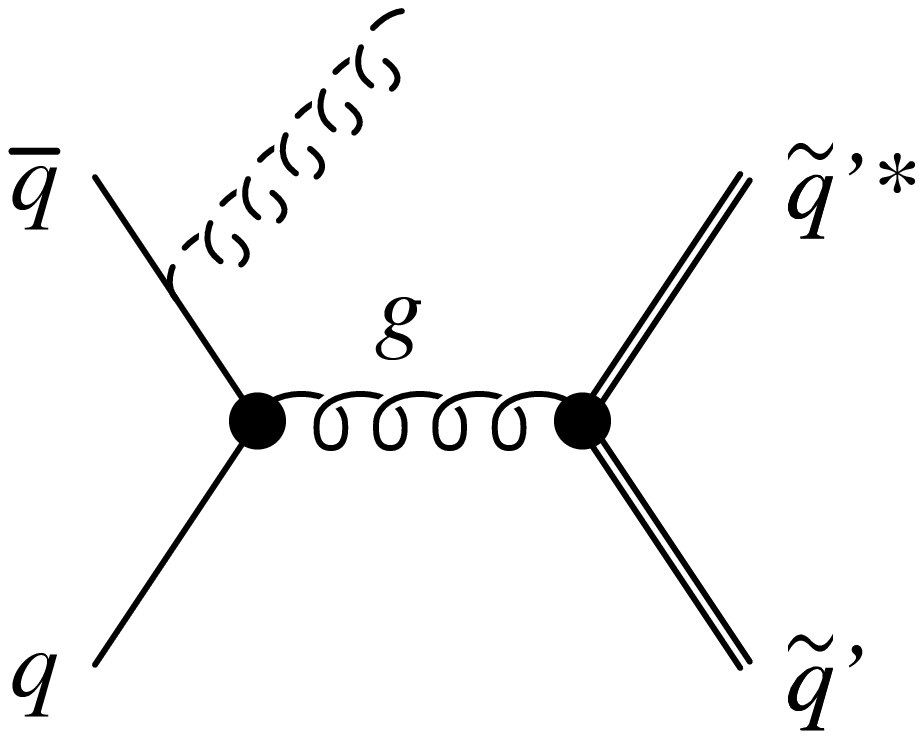, height=3.5cm}&
\psfig{figure=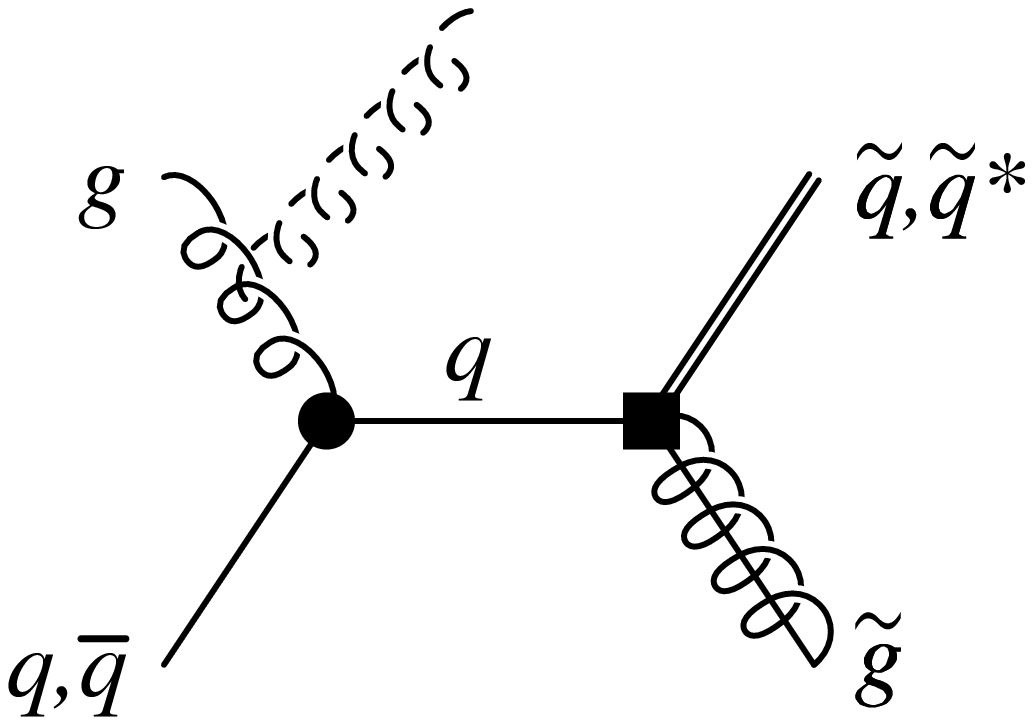, height=3.5cm}&
\psfig{figure=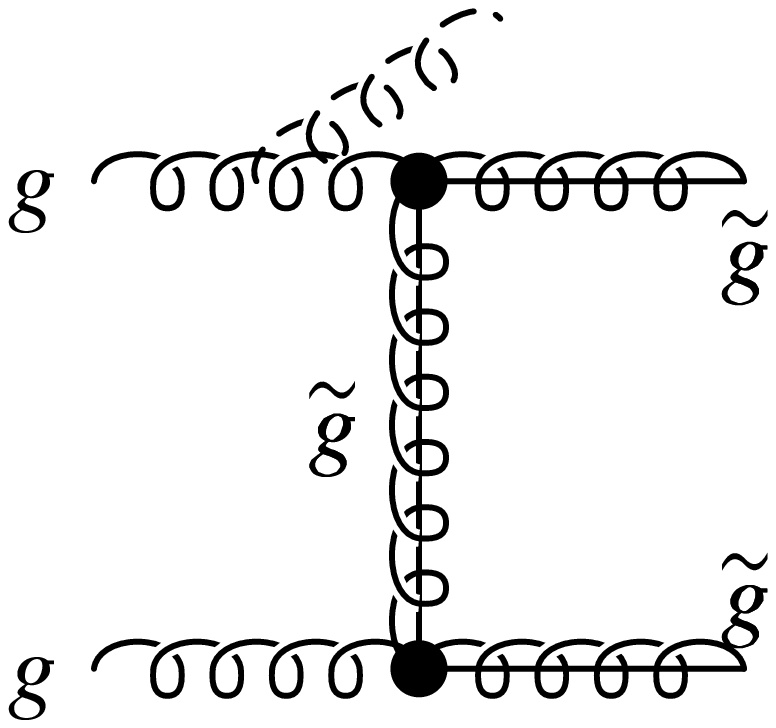, height=3.5cm}
\end{tabular}
\mycaption{Characteristic examples of Feynman diagrams for
partonic squark and gluino production in hadron
collisions. Dots indicate the gauge coupling $g_{\rm s}$, while squares represent
for the Yukawa coupling $\hat{g}_{\rm s}$.
Gluon radiation in the initial state, indicated by the dashed lines, can lead to
extra hard jets.}
\label{fg:dia1}
\end{figure}

Most suited for this purpose, in the area
of parameter space in which gluinos are heavier than squarks, are
the basic subprocesses
\begin{eqnarray}
 \mbox{squark-pair production} &:& q \,q \to \tilde{q} \,\tilde{q} \,(g) 
 \label{eq:qq} \\
 \mbox{Compton process}        &:& g \,q \to \tilde{q} \,\tilde{g} \,(g)
                                   \;\, {\rm with} \;\,
                            \tilde{g} \to \tilde{q} \,q                      
  \label{eq:qg}
\end{eqnarray}
{\it cf.} Fig.~\ref{fg:dia1}. Gluinos decay instantly to squarks under this
kinematical  condition. $(g)$ denotes additional gluon jets emitted in the
production  process. The analysis of this kinematical area, admittedly, is
easier than the reverse mass configuration but sufficient to illustrate the
basic points. Other processes of secondary importance, like gluino-gluino
pair production {\it etc.}, have been included in the final analysis properly. 
As evident from Fig.~\ref{fg:dia1}, the size of the cross section for
squark-pair  production is set by the fourth power of the strong Yukawa
coupling, $\hat{g}^4_{\rm s}$, while the cross section of the Compton process is 
determined by the product of the second power of the gauge and Yukawa 
couplings, $g^2_{\rm s} \hat{g}^2_{\rm s}$. Since the relative weight of the 
subprocesses changes for different jet final-state configurations, the  Yukawa
coupling can be measured by comparing the cross sections in  different parts 
of the phase space.  

To focus on the basic processes Eqs.~\eqref{eq:qq},\eqref{eq:qg} 
and to reduce the backgrounds
from  other SUSY processes, squark-squark pairs of the same charge 
should be 
selected. This can be achieved by studying like-sign di-lepton final states
\cite{quigleys} in gaugino-type chargino and neutralino decay chains of two L- and R-type 
squarks like
\begin{gather}
\begin{array}{rclcl}
 \tilde{u}_{\rm L}   &\to& d + \tilde{\chi}^+_1 &\to& \text{jet} + \ell^+ + \Eslash            \\[.5mm]
 \tilde{u}_{\rm L,R} &\to& u + \tilde{\chi}^0_2 &\to& \text{jet} + \ell^+/\ell^- + \Eslash 
\end{array}
		     \\[1.5mm]
\begin{array}{rclcl}
 \tilde{d}_{\rm L}   &\to& u + \tilde{\chi}^-_1 &\to& \text{jet} + \ell^- + \Eslash            \\[.5mm]
 \tilde{d}_{\rm L,R} &\to& d + \tilde{\chi}^0_2 &\to& \text{jet} + \ell^-/\ell^+ + \Eslash  
\end{array}
\end{gather}
with $\Eslash$ denoting the missing energy. 
SU(2) isospin invariance guarantees 
that the partial 
widths for $\tilde{u}_{\rm L}$ and $\tilde{d}_{\rm L}$ decays to charginos 
are equal to very good approximation in this configuration. 
Assuming CP-invariance, the subsequent leptonic decays
of $\tilde{\chi}_1^-$ are related to $\tilde{\chi}_1^+$, and the $\tilde{q}_{\rm L}^\ast$ 
decays to the $\tilde{q}_{\rm L}$ decays. For pure SU(2) singlet gaugino LSP
states $\chi^0_1$, R-type squarks $\tilde{q}_R$ will decay into the channel 
$q_R + \chi^0_1$, not generating charged leptons in the final state.

Extended mixing in the third-generation stop and sbottom sector will
significantly increase the complexity of the analysis [though the production of
stops may prove particularly useful in testing the Majorana character of
gluinos \cite{Maj}]. However, scalar top and bottom decays generate $b$ quarks
in the final states which, by tagging, can be used  to strongly reduce the
influence of these sparticles on the analysis. 

The branching ratios for gluino decays to squarks plus quarks depend only on 
the gluino, squark and quark masses 
so that they can be predicted unambiguously in the Compton process. 
To leading order the widths are given by the Yukawa couplings which 
however cancel from the branching ratios.

\section{Di-lepton plus jet analyses in the MSSM}
\label{dMSSM}

In the first step we have elaborated on the details of the scheme 
outlined above within the SUSY scenario defined in Ref.~\cite{quigleys}. Apart 
from the slightly increased gluino mass to improve on jet tagging, the scenario 
corresponds to the reference point SPS1a \cite{sps},
describing a specific light-mass scenario within the minimal supersymmetric
standard model MSSM.\footnote{The model in Ref.~\cite{quigleys} is defined
strictly at the weak scale and is not intended to accurately reflect either GUT
scale physics or cosmological parameters which most likely are affected by other
mechanisms too.}
Typical parameters, masses 
and branching ratios relevant to the present analysis, are collected 
in Table~\ref{tab:scen}. 
\begin{table}[tb]
\hrulefill\\
a) {\it Weak scale soft supersymmetry breaking parameters:}
$$
\begin{aligned}
M_1 &= 101 \gev & \quad m_{\rm L3} &= 199 \gev & \qquad m_{\rm Q1} &
= 551 \gev & \quad m_{\rm Q3} &= 499 \gev \\
M_2 &= 192 \gev& m_{\rm R3} &= 136 \gev & m_{\rm U1} &= 529 \gev &
m_{\rm U3} &= 418 \gev\\
\mu &= 352 \gev& & & m_{\rm D1} &= 526 \gev &
m_{\rm D3} &= 523 \gev\\
\tan\beta &= 10 & A_\tau &= -256 \gev & A_{\rm b} &= -797 \gev & A_{\rm b} &= -505 \gev   \\[-1em]
\end{aligned}
$$
\hrulefill\\
b) {\it Superpartner pole masses:}
$$
\begin{aligned}
m_{\suL} &= 567 \gev & \quad m_{\sdL} &= 573 \gev
	& \quad m_{\suR} &= 547 \gev & m_{\sdR} &= 546 \gev\\
m_{\tilde{t}_1} &= 396 \gev & m_{\tilde{b}_1} &= 515 \gev
	& m_{\tilde{t}_2} &= 586 \gev & m_{\tilde{b}_2} &= 544 \gev\\
\mneu{1} &= 97 \gev & \mneu{2} &= 181 \gev
	& \mcha{1} &= 176 \gev & \quad \mneu{3,4},\mcha{2} &\sim 370 \gev \\
m_{\st_1} &= 136 \gev & m_{\st_2} &= 208 \gev & m_{\go} &= 700 \gev  \\[-1em]
\end{aligned}
$$
\hrulefill\\
c) {\it Relevant branching ratios:}
$$
\begin{aligned}
\text{BR}[\suL \to d \, \cha^+_1] &= 65\% &\quad
 \text{BR}[\cha^\pm_1 \to \tilde{\tau}^\pm \nu_\tau] &= 95\% &\quad
 \text{BR}[\go \to q\, \sqL] &=  \phantom{0}7.1\%, & q\neq b,t \\
\text{BR}[\sdL \to u \, \cha^-_1] &= 61\% &\quad
 \text{BR}[\cha^\pm_1 \to W^{*\pm} \neu_1] &= \phantom{0}5\% &
 \text{BR}[\go \to q\, \sqR] &=  \phantom{0}9.6\%, & q\neq b,t \\
\text{BR}[\suL \to u \, \neu_2] &= 32\%
&&&
 \text{BR}[\go \to b \, \tilde{b}_i] &= 23\% \\
\text{BR}[\sdL \to d \, \neu_2] &= 31\%
&&&
 \text{BR}[\go \to t \, \tilde{t}_1] &= 10\% \\[-1em]
\end{aligned}
$$
\hrulefill
\mycaption{Reference scenario used in this analysis.
It coincides with the MSSM Snowmass point SPS1a \cite{sps},
except for the slightly larger gluino mass \cite{quigleys}.
The low-energy spectrum has been calculated by means of SoftSUSY 1.9.1
\cite{softsusy}.}
\label{tab:scen}
\end{table}

Similarly to Ref.~\cite{quigleys},
we will assume in this section that all the parameters, masses and 
branching ratios, have been pre-determined in squark cascades at LHC and/or
analyses at an $e^+e^-$ linear collider. It should be noted, however, that within the
group-theoretical frame of the minimal supersymmetric extension of the
Standard Model [MSSM], the squark branching ratios can be predicted if 
the structure of the chargino/neutralino sector has been explored at a
low-energy lepton collider. Within this framework the measurement of the 
squark-decay branching ratios can be substituted by the completeness 
assumption of the MSSM. The
assumption can be cross-checked internally at LHC by searching for non-MSSM 
squark decays. 

About two thirds of the time, the light-flavor L-squarks decay into 
the charginos $\tilde{\chi}^\pm_1$, 
while about one third of the decays go into the neutralino $\neu_2$,
following from the fact that the L-squarks have small hypercharges, $Y_{\sqL} =
1/6$, and thus
predominantly couple to winos. Furthermore, since the mixing between the bino and
wino components of the neutralinos is small, as well as the mixing between L-
and R-squark states negligible, the branching fractions of the up- and
down-squarks are almost equal. This is in particular the case in our
reference scenario, {\it cf.}~Table~\ref{tab:scen}, and a consequence of 
SU(2) and CP invariance. [Note that in general CP invariance can be broken in the MSSM
by complex parameters, but in this analysis all supersymmetry and supersymmetry
breaking parameters are assumed to be real.]

Due to the relatively large value of $\tan\beta = 10$ in this scenario, the
charginos $\cha^\pm_1$ almost exclusively decay to the scalar tau $\tilde{\tau}_1$, 
which subsequently decays
to taus. To trace the charge of the squarks explicitly, we restrict ourselves
to the leptonic tau decays. The phenomenological analysis we
will therefore be designed such that the decay chains
\begin{eqnarray}
\suL & \stackrel{65\%}{-\!\!\!-\!\!\!\longrightarrow} & d \, \cha^+_1
\stackrel{100\%}{-\!\!\!-\!\!\!\longrightarrow} d \, \tau^+ \, \nu_\tau \,
\neu_1 \stackrel{35\%}{-\!\!\!-\!\!\!\longrightarrow} d \, \ell^+ + \Eslash \nonumber\\
\sdL & \stackrel{61\%}{-\!\!\!-\!\!\!\longrightarrow} & u \, \cha^-_1
\stackrel{100\%}{-\!\!\!-\!\!\!\longrightarrow} u \, \tau^- \, \bar{\nu}_\tau \,
\neu_1 \stackrel{35\%}{-\!\!\!-\!\!\!\longrightarrow} u \, \ell^- + \Eslash \;\;
                                                                    [\ell = e, \, \mu]
\label{eq:dec}
\end{eqnarray}
are singled out.
The numbers above the arrows denote the branching fractions.

The LHC processes which have been studied in this report are like-sign di-lepton
final states plus a number of jets with variable transverse momenta:
\begin{eqnarray}
 pp \to \ell^\pm \ell^\pm + 2 \text{ jets} + \Eslash     \label{eq:2j}       \\
 pp \to \ell^\pm \ell^\pm + 3 \text{ jets} + \Eslash     \label{eq:3j}       \\
 pp \to \ell^\pm \ell^\pm + 4 \text{ jets} + \Eslash     \label{eq:4j}
\end{eqnarray}
and possibly additional jets from gluon radiation, that however tend to have smaller 
transverse energies than the jets emitted in squark and gluino decays.

The first process, Eq.$\,$\eqref{eq:2j}, concentrates on direct squark-pair production,
the classical  reaction~\cite{quigleys} for measurements of the strong Yukawa
coupling with the basic cross section scaling as $\hat{g}^4_{\rm s}$. The
transverse momentum of the jets is assumed beyond 200 GeV.

In the second process, Eq.$\,$\eqref{eq:3j}, we demand two high transverse
momentum jets with $E_{\rm T} \geq 200$ GeV, plus an additional third jet with variable
transverse momentum.  For low transverse momenta of the third jet, direct
squark-pair production is projected out, moving the dominant weight to the
Compton process however for increasing transverse momentum. In this way, 
depending on the transverse momentum,
different powers of the Yukawa  coupling, 
$\sim c_1(p_{\rm T}) \hat{g}^4_{\rm s} + c_2(p_{\rm T}) g^2_{\rm s} \hat{g}^2_{\rm s}$, 
become relevant. 

The evaluation of the third process, Eq.~\eqref{eq:4j}, parallels the previous
analysis. It does not provide new methodological insight but increases
statistics when included.

\paragraph{ }
To analyse the prospects for extracting the SUSY-QCD Yukawa coupling in a
realistic experimental environment, we have generated samples for the squark
and gluino production process as well as SM backgrounds with {\sc Pythia} 6.408
\cite{pythia}, including gluon and quark radiation through virtuality-ordered
parton showers, string hadronisation and the underlying event according to
`Tune A' \cite{tunea}. The required modifications for varying the SUSY-QCD
Yukawa coupling independently of the gauge coupling were implemented
in a private  version\footnote{Publicly available from {\sc
    Pythia} version 6.4.19.},  
according to the lowest-order cross sections given in \cite{Hop}.
The main SM backgrounds originate from $t\bar{t}$
production as well as $WWjj$ production. The $WWjj$ process was generated with
{\sc MadGraph} \cite{mad}, without QCD radiation and fragmentation.

The cross-sections for squark, gluino and top production were
normalised by $K$-factors to the values obtained with next-to-leading order
QCD corrections~\cite{Hop,Ellis}, while for the $W^\pm
W^\pm jj$ background only leading order results are available. 
In analysing distributions, K-factors as calculated for the total cross-sections have been
included; this approximation is based on conclusions of Ref.~\cite{Hop} where the primary distributions
have been shown not to be distorted by more than 10\% so that a uniform rescaling is justified.
A possible complication is created by the fact that 
the gluino is quite wide in our scenario, with
$\Gamma_{\tilde{g}}\sim 35\ $GeV at
leading order, generating doubts whether the narrow-width
limit is justified. For example, the convolution of the 
corresponding Breit-Wigner function(s) with the steeply falling parton
distributions greatly distorts the final distribution -- to the
point of even creating a spurious peaking towards low masses. We note
that this issue is generic for gluon-induced processes involving wide
resonances at the LHC, and although we do not aim to solve it here,
the reader should be aware of this problem.  

Finally, the detector response was parametrised by simple acceptance cuts:
a toy calorimeter
spanning the pseudorapidity range $|\eta|<5$ with a
resolution of $0.1\times 0.1$ in $\eta\times\phi$ space was assumed. 
Jets were identified with a simple UA1-like cone jet algorithm (\texttt{PYCELL}),
with a cone size of $\Delta R=0.4$. Muons and electrons
were reconstructed inside $|\eta|<2.5$ and an
isolation criterion was further imposed, requiring both less
than 10 GeV of additional energy deposited in a cone of size
$\Delta R=0.2$ around the lepton and also no reconstructed jets
with $p_{\rm T,j} > 25$ GeV closer than $\Delta R=0.4$ around the
lepton. These criteria duplicate the default settings of the ATLFAST simulation
package \cite{atlas}.

All the signal processes, Eqs.~\eqref{eq:2j} to~\eqref{eq:4j}, are characterised
by the similar features of two leptons, missing energy and at least two hard
jets. Therefore the following preselection cuts have been used in the event
generation:
\begin{itemize}
\item At least $\Eslash > 100$ GeV of missing energy.
\item At least 2 jets with $p_{\rm T,j} > 100$ GeV.
\item Exactly two isolated same-sign
leptons $\ell = e,\mu$ with $p_{\rm T,\ell}  > 7$ GeV.
\end{itemize}
To reduce the SM background, the final analysis in addition employs the
soft selection cuts
\begin{itemize}
\item At least $\Eslash > 300$ GeV of missing energy.
\item At least 2 jets with $p_{\rm T,j} > 200$ GeV.
\item Veto on identified bottom-quark jets, using a tagging efficiency of
$\epsilon = 90$\% with a mistagging rate $D = 25$\% \cite{atlas,quigleys}.
\end{itemize}
These cuts have been optimised to minimise the remaining SM background, however
at the cost of being relatively expensive on the signal statistics as well.
This choice is justified by the goal to interpret the measured signal rate
and distributions as a combination of $\tilde{u},\tilde{d}$
squark and $\tilde{g}$ gluino production processes, so that it is mandatory 
to keep the contamination from background processes as low as possible.

The veto on bottom quarks is effective against the $t\bar{t}$ background, but
it also helps to reject the unwanted decay modes of the gluinos into scalar
bottoms and scalar tops. Since the stop and sbottom decay chains can deviate
significantly from those of the light-flavour squarks, it is advantageous to
remove them from the analysis. As mentioned above, the relevant branching
ratios of the gluinos into different squark flavours can be determined
unambiguously from the gluino, squark and quark masses.

After application of the cuts, the
signal and background cross-sections listed in Tab.~\ref{tab:xsec} are obtained.
\begin{table}
\renewcommand{\arraystretch}{1.35}
\begin{tabular}{|l||r|r||r|}
\hline
Process & Total cross-section & Cross-section with BRs and cuts & Efficiency \\
\hline\hline
$\sigma[\sqL\sqL]$ & 2.1 pb & 6.1 fb & 0.29\% \\
$\sigma[\sqL\sqL^*]$ & 1.4 pb & 3.1 fb & 0.23\% \\
$\sigma[\sqL\go]$ & 7.0 pb & 7.6 fb & 0.11\% \\
$\sigma[\go\go]$ & 3.2 pb & 1.4 fb & 0.04\% \\
\hline
$\sigma$[SM] & 800 pb & $<$0.6 fb & \\
\hline
\end{tabular}
\mycaption{Signal and background cross-sections after applying the cuts
mentioned in the text.}
\label{tab:xsec}
\end{table}
Thus, with a total luminosity of 300 fb$^{-1}$ about 2700 squark pair events and
1750 events from the Compton process are expected.

\begin{figure}[tb] \centering 
\epsfig{figure=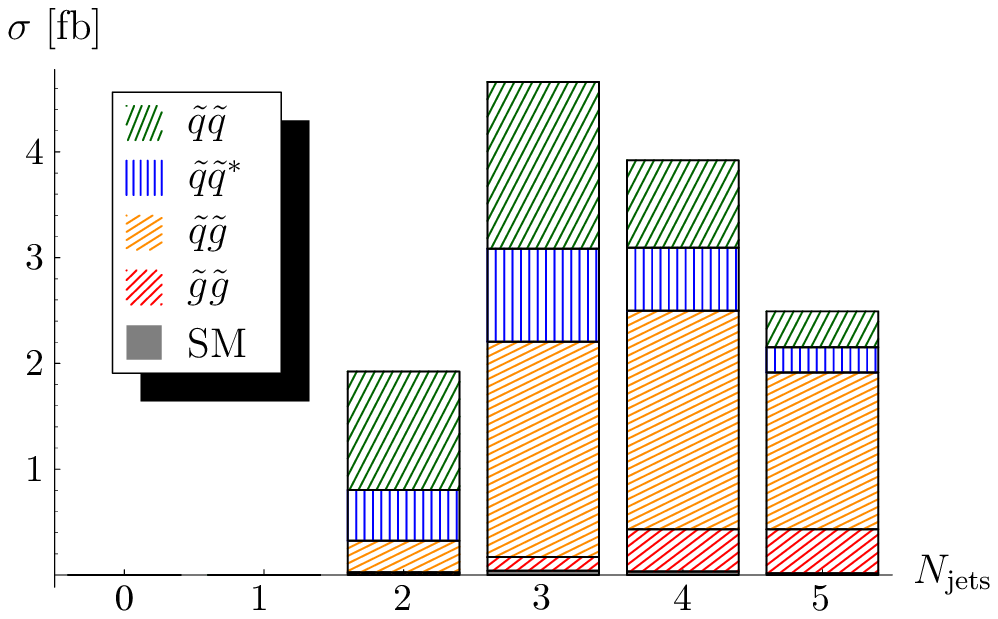, width=12cm, bb=25 490 363 662}
\mycaption{Results of simulations for the number of
identifiable jets, for the different
processes that lead to the general signatures in Eqs.~\eqref{eq:2j}
to~\eqref{eq:4j}.
For the first two jets, a threshold of $p_{\rm T,jet} > 200$
GeV has been used, while $p_{\rm T,jet} > 50$ GeV for the other jets.
} \label{fg:nj} \end{figure}

\paragraph{ }
After reducing the SM background, the three high-energy
processes $pp \to \sqL\sqL$, $\sqL\go$, $\go\go$ should be
distinguished in order to extract the SUSY-QCD Yukawa coupling
$\hat{g}_{\rm s}$. As outlined above, in a first step this can be achieved, statistically, 
by analysing the number of hard jets in the final state, see Fig.~\ref{fg:nj}.
With a transverse momentum threshold of $p_{\rm T,jet} > 50$ GeV for
the jets beyond the two hardest jets (with $p_{\rm
T,jet} > 200$ GeV), the sample of events with only two jets is dominated by
squark
pair production.  On the other hand, events with three or more jets
largely originate from the Compton process $pp \to\sqL\go$. For $N_{\rm
jets} \ge 4$, gluino pair production also contributes to the di-lepton $+$
jets signal, though the rate for this process is strongly suppressed 
due to the $b$ veto.

From the measured rates for events with 2, 3 and 4 jets, and using the
pre-determined branching ratios for squarks and gluinos, the coupling
$\hat{g}_{\rm s}$ can be determined. A similar analysis has already been
performed in Ref.~\cite{quigleys}, but only using the 2-jet portion. Including
the higher statistics of the 3- and 4-jet events, we  conclude that a much more
precise measurement of $\hat{g}_{\rm s}$ could be possible. From a fit to the
simulated data, the statistical error is estimated to be only 0.4\%, assuming a
total luminosity of 300 fb$^{-1}$.

With this  precision, systematic errors might become the limiting factor for
the accurate determination of $\hat{g}_{\rm s}$. We have identified and
estimated the following dominant sources of systematic uncertainties.

The extraction of a cross section for a hard production process at a hadron collider depends on the knowledge of the parton distribution functions (PDFs) of the protons in the initial state. It is likely that our knowledge of PDFs for processes with large transverse momentum will be improved by data taken at the LHC itself. 
Nevertheless, here we adopt a conservative approach 
and estimate the uncertainty resulting from PDFs by comparing results for different
CTEQ PDFs releases (CTEQ6M,  CTEQ6D,  CTEQ5M1, CTEQ5L and CTEQ6L1) \cite{cteq} that are
available today. It is found that this variation leads to an uncertainty 
of about 1\% in the extracted Yukawa coupling.

Furthermore, the precise analysis of the squark and gluino cross sections
relies on accurate squark and gluino mass measurements and on theoretical
calculations of higher-order corrections. For the sparticle mass errors, we
have adopted the results of Ref.~\cite{lhclc}, where an uncertainty of about 10~GeV
is expected for the L-squark masses and about 12~GeV for the gluino masses. On
the theoretical side, the next-to-leading order corrections are known
\cite{Hop} and have been included in the analysis. The uncertainty of the
missing ${\cal O}(\alpha^2_{\rm s})$ contributions are estimated by varying the
renormalisation scale of the ${\cal O}(\alpha_{\rm s})$ corrected cross section
between $m/2 < Q < 2m$, where $m$ is the average mass of the produced
sparticles (squark or gluino).

The error estimate does not include the uncertainty for the squark
branching ratios which are presumed to be small. In practice, due to the
large squark masses in our reference scenario, they could either
be measured at an $e^+e^-$ linear collider with energy beyond 1 TeV or,
based alternatively on the closure of the MSSM, one could reconstruct
the neutralino/chargino sector at a 500 GeV linear collider
with high precision at the per-cent to per-mil level \cite{CKMZ,SFIT}
and derive the squark branching fractions from this information.

The effect of the systematic errors and the combined error, obtained by
adding all individual error sources in quadrature, are listed in
Table~\ref{tab:err}~(Method A).  It turns out that the total error, $\delta
[\hat{g}_{\rm s}/g_{\rm s}] = 3.4\%$ is dominated by systematic effects.

With a luminosity upgrade of the LHC, yielding a total luminosity of 1 ab$^{-1}$, 
and assuming that the mass and PDF uncertainties will improve with
higher luminosity, this error could be reduced to 2.5\%.
\begin{table}
\centering
\renewcommand{\arraystretch}{1.35}
\begin{tabular}{|l|rrr|}
\hline
Error source & \multicolumn{3}{c|}{$\delta[\hat{g}_{\rm s}/g_{\rm s}]$}
              \\
\hline
\multicolumn{1}{|r|}{Method} & A & B & C \\
\hline
LHC statistics for 300 fb$^{-1}$ & 0.6\% & 3.0\% & 4.9\% 
\\
\hline
PDF uncertainty & 1.4\% & 3.0\% & 3.0\%  \\
NNLO corrections & 2.0\% & 3.2\% & 3.2\%  \\
Mass measurements $\Delta m_{\sq} = 10$ GeV & 2.0\% & 1.2\% & 1.2\% \\
\phantom{Mass measurements}
$\Delta m_{\go} = 12$ GeV & 1.1\% & 2.2\% & 2.2\%  \\
\hline
Total & 3.4\% & 5.9\% & 7.3\%  \\
\hline
\end{tabular}
\mycaption{Combination of statistical and systematic errors for extracting
the SUSY-QCD Yukawa coupling from like-sign dilepton events including 
jet transverse-momentum distributions at the LHC.
In {\rm Method A} the complete information on squark branching ratios
from a multi-TeV $e^+e^-$ collider
is used for an SPS1a-type scenario.
{\rm Method B} is based on LHC data potentially 
alone, but supported eventually by 1 TeV $e^+e^-$ collider data, and combined with 
moderate assumptions on new heavy iso-scalar singlets. {\rm Method C} allows  
for more complex scenarios in the neutralino sector than encoded
in SPS1a-type reference points, allowing again for the presence of new Standard
Model singlets.}
\label{tab:err}
\end{table}

\section{Towards model-independent analyses}
\label{iMSSM}

The procedure of determining the Yukawa coupling $\hat{g}_{\rm s}$ 
as outlined in the previous section can be expanded
towards a more general analysis. Though the problem is too complex to allow
for a completely model-independent analysis, methods can nevertheless be
exploited which are based only on the pre-determined 
chargino/neutralino system, as expected to be known from LHC and/or
$e^+e^-$ collider measurements, combined with natural assumptions on the structure
of theories beyond the minimal supersymmetric extension of the Standard Model
MSSM. Central to this approach is the circumvention of external
measurements of heavy squark-decay branching ratios which, for general expectations
of the spectrum, can only be performed at a multi-TeV lepton collider.

\subsection{MSSM plus heavy iso-singlets} 

First we will analyse a scenario in which the MSSM is expanded by a set of 
heavy iso-scalar particles, with little mixing between standard and new
states, as could be expected in extended Higgs and gauge theories, see
{\it e.g.} Refs.$\,$\cite{HiU}.

\paragraph{ } 
Proceeding from the same scenario as analysed in section \ref{dMSSM}, 
we observe that the charginos and neutralinos are light enough to be produced 
as diagonal and mixed pairs in $e^+e^-$ collisions below 1 TeV, 
${\tilde{\chi}^+_i} {\tilde{\chi}^-_j}$ $(i,j=1,2)$
and ${\tilde{\chi}^0_i} {\tilde{\chi}^0_j}$ $(i,j=1,\ldots,4)$.
Masses and mixing parameters can be measured in $e^+e^-$ experiments,
{\it cf.} Refs.$\,$\cite{CKMZ}, possibly even by observing only a subgroup
of the lightest particles \cite{D}. They determine the SU(2), U(1) gaugino and
higgsino decomposition of the charginos and neutralinos. 
Thus the partial widths $\Gamma(\tilde{q} \to \tilde{\chi})$ can be predicted 
from these experimental results.
Moreover, the decay
branching ratios $BR(\tilde{\chi} \to \ell^\pm)$ can be measured directly.  
In many scenarios even the observation of relatively large mass differences among 
the neutralinos and charginos [possible at LHC already] is sufficient to suggest
strongly that
the mixing in the neutralino and chargino sector is very small and thus can be neglected 
for the purposes here. This is the case, for instance, in the reference scenario studied 
in this paper. 

While the partial widths $\Gamma(\tilde{q} \to \tilde{\chi} \to \ell)$ can be 
completely determined in this scenario, the total widths $\Gamma(\tilde{q})$,
and as a result the branching ratios, remain unknown since decay channels to 
heavy particles beyond MSSM, in particular to heavy iso-singlets, may be open. However, 
since R-squarks do not decay to leptons in scenarios where the lightest neutralino is predominantly bino, 
the only unknown quantities are the branching ratios
$BR(\tilde{u}_L \to \tilde{\chi}^+)$ and $BR(\tilde{d}_L \to \tilde{\chi}^-)$
to which all other branching fractions can be linked once the $\tilde{\chi}$ mixing 
matrices
are known; according to the Wigner-Eckart theorem:
\begin{eqnarray}
   BR(\tilde{u}_L \to \tilde{\chi}^0_2) &=& \frac{1}{2} \; f_{uu}(N^\pm,N^0) \, 
                                            BR(\tilde{u}_L \to \tilde{\chi}^+_1) \nonumber \\
   BR(\tilde{d}_L \to \tilde{\chi}^0_2) &=& \frac{1}{2} \; f_{dd}(N^\pm,N^0) \,
                                            BR(\tilde{d}_L \to \tilde{\chi}^-_1) 
\end{eqnarray}
The coefficients $f_{uu,dd}$ are ratios of the chargino $N^\pm$ and the neutralino $N^0$  
mixing matrix elements, with $\frac{1}{2}$ being the ratio of the standard 
Clebsch-Gordan coefficients
squared. For pure $\tilde{\chi}^\pm_1, \tilde{\chi}^0_2$ iso-vector winos, 
the relations simplify to $f_{uu,dd} \to 1$.

This could be the platform for a 3-parameter analysis, including the unknown
branching ratios and the strong Yukawa coupling. 
However, the two branching ratios above are related 
if squark decays beyond $\tilde{\chi}^\pm_i$ and $\tilde{\chi}^0_i$ 
involve only new iso-singlet particles with universal couplings to L-squarks,
as may be expected, on
quite natural grounds, outside the Standard Model. In the same notation as above, the
relation reads
\begin{equation}
   BR(\tilde{d}_L \to \tilde{\chi}^-_1) = f_{du}(N^\pm,N^0) \,
                                          BR(\tilde{u}_L \to \tilde{\chi}^+_1)
\end{equation}
[with $f_{du} \to 1$ in the wino limit]. In this scenario the analysis
of section \ref{dMSSM} repeats itself except for leaving free the overall 
normalisation, governed by the unknown branching ratio $BR(\tilde{u}_L \to \tilde{\chi}^+_1)$. 
In this case, the SUSY-QCD Yukawa coupling can only be 
extracted from ratios of the cross sections or from distributions with varying transverse 
momenta, but not from absolute values of cross sections.  

By analysing the measured rates for events with 2, 3 and 4 jets, one can
extract a ratio of the cross-sections for $\sqL\sqL$ and $\sqL\go$ production.
This  has also been exploited for the model analyses in Ref.~\cite{noj07}. By
comparison with the theoretical calculations for these cross-sections, this
ratio can then be interpreted in terms of the coupling ratio $\hat{g}_{\rm
s}/g_{\rm s}$.

\paragraph{ } 
The robustness and precision of the Yukawa coupling determination can be
improved by looking not only at the number of identified jets, but also their
transverse momentum distribution. 

\begin{figure}[tbp]
\raisebox{7cm}{(a)}
\epsfig{figure=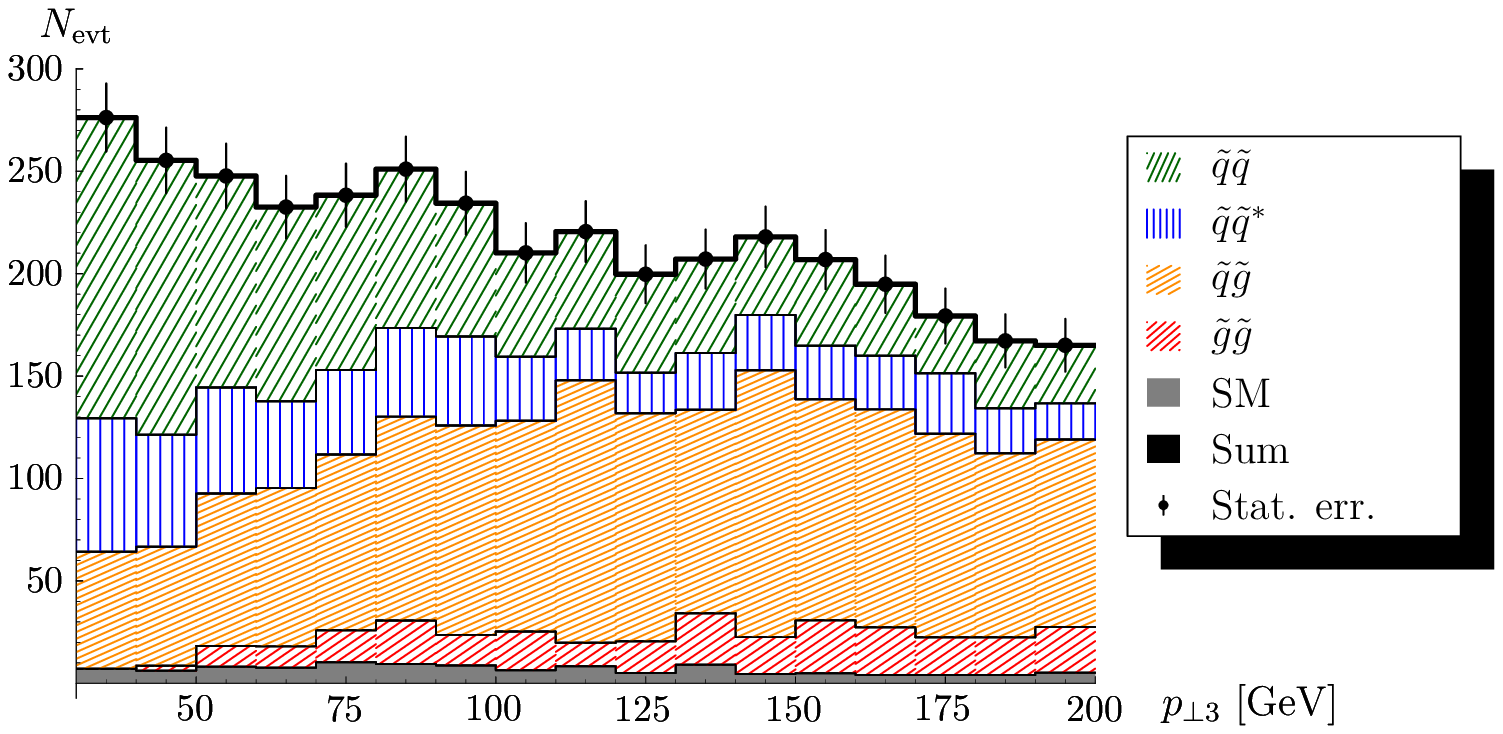, width=15cm, bb=30 450 460 660} \\[1em]
\raisebox{7cm}{(b)}
\epsfig{figure=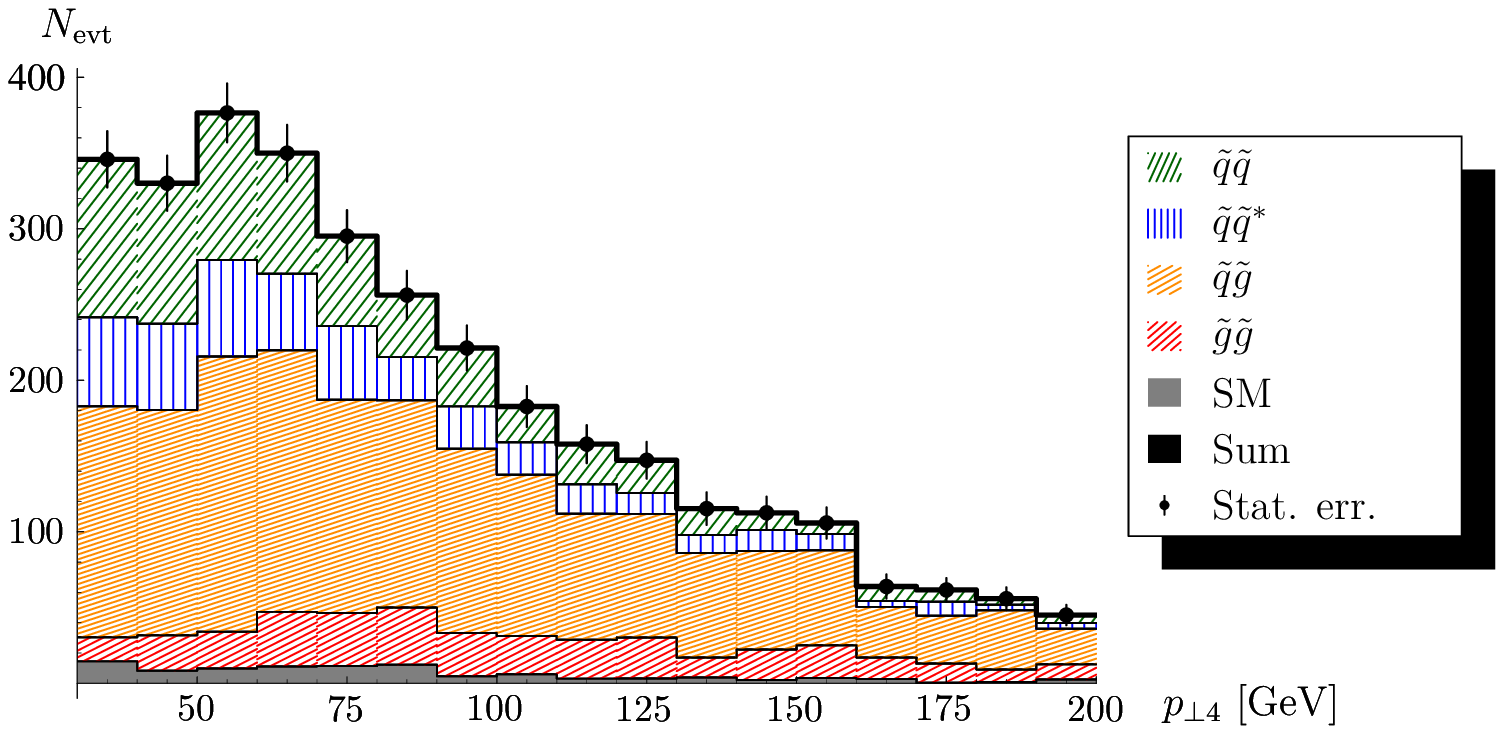, width=15cm, bb=30 450 460 660}
\mycaption{Results of simulations for (a) the distribution with respect to the
$p_{\rm T}$ of the third jet, and (b) the distribution with respect to the
$p_{\rm T}$ of the fourth jet (with $p_{\rm T,3} > 50$ GeV). In both plots the
contributions from the processes in Eqs.~\eqref{eq:2j} to~\eqref{eq:4j} as well
as the SM background are
shown separately, while for the sum the statistical errors for 300 fb$^{-1}$
is also shown.}
\label{fg:res}
\end{figure}

The results of the simulations are presented in 
Figs.$\,$\ref{fg:res}.  
As apparent from the figure, both the distributions as a function of the $p_{\rm T}$ 
of the third and the fourth jet can discriminate between the two processes
Eqs.$\,$\eqref{eq:qq} and \eqref{eq:qg}. Squark-pair production tends to
peak towards low values of $p_{\rm T,3}$ and $p_{\rm T,4}$, since the
additional jets are generated in this process 
only by gluon radiation from the initial or final
state. The $\tilde{q}\tilde{g}$ Compton production and the $\go\go$ gluino-pair
production, on the other hand, typically lead to larger values for $p_{\rm T,3}$
and $p_{\rm T,4}$. Therefore, the lower ends of the $p_{\rm T,3}$ or $p_{\rm
T,4}$ spectra depend more strongly on variations of $\hat{g}_{\rm s}$ than the
region of higher transverse momenta. As a consequence, the shape of the
transverse momentum distributions is sensitive to the ratio $\hat{g}_{\rm
s}/g_{\rm s}$, independent of the total normalisation of the cross-section. 

The expected experimental
precision for the determination of the SUSY-QCD Yukawa coupling can be
estimated from a binned $\chi^2$ fit. Assuming 300 fb$^{-1}$ of integrated
luminosity, the statistical error for $\hat{g}_{\rm
s}/g_{\rm s}$ is 3.3\%. 
This has to be combined with systematic errors as given in Table~\ref{tab:err}~(Method B),
generating a final error of 5.9\%. 
The systematic errors were estimated in the same way as explained in the previous 
section.

\begin{figure}[tb]
\centering
\epsfig{figure=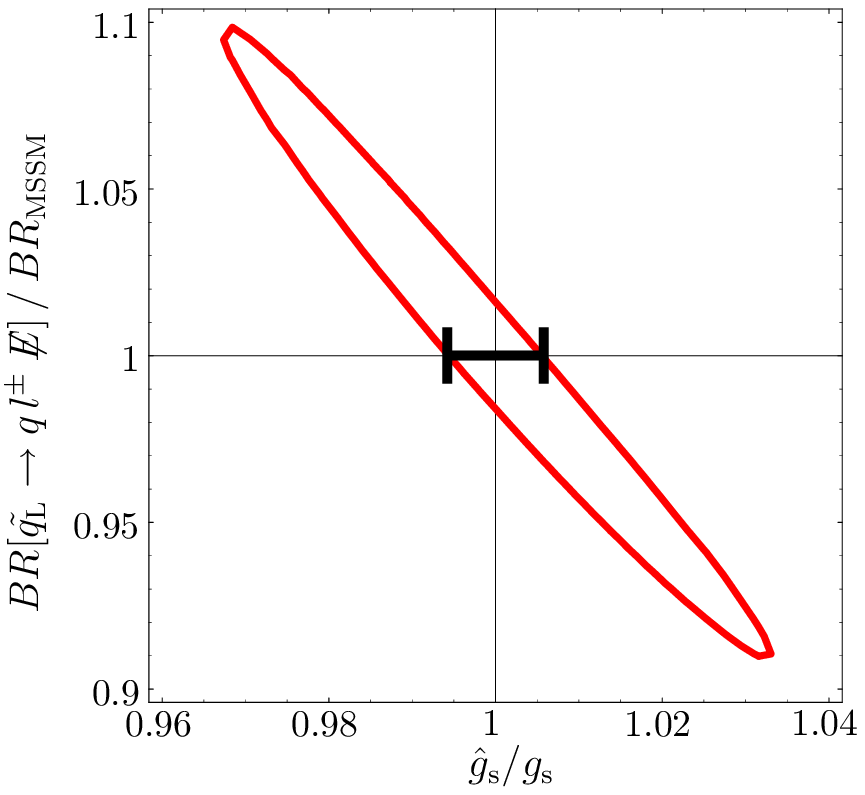, width=8cm}
\mycaption{Independent results for the SUSY-QCD Yukawa coupling $\hat{g}_{\rm
s}$ and the overall branching ratio of L-squarks into a single identified lepton 
$l = e,\mu$ (normalised to the MSSM value)
from a fit to the $p_{\rm T3}$ distribution  in 
Fig.~\ref{fg:res}~(a). The plot shows the 1$\sigma$ statistical error
contours corresponding to the fit
method B. The black bar indicates the result from the fit method A, assuming that the
branching ratio has been pre-determined.}
\label{fg:bryuk}
\end{figure}
Since the overall normalisation is left as a free parameter in this analysis, the 
result can also be exploited to determine the branching ratio of squark decays to 
charginos. A fit to $p_{\rm T,3}$, for instance, can then simultaneously be 
interpreted as information about the Yukawa coupling and the
squark branching ratio, see Fig.~\ref{fg:bryuk}.

\subsection{Generalised picture}

In the second more general step we drop the assumption that 
$\tilde{\chi}^0_2$ is nearly wino-like but allow, as before, for a set 
of new heavy singlets under the Standard Model group. 

In this scenario, the neutralino sector becomes more involved. Significant
mixing between bino and wino components in the neutralino $\tilde{\chi}^0_2$,
for instance, 
can create a sizable branching ratio of R-squarks
$\tilde{u}_R,\tilde{d}_R$ into leptons [considering, as before, the decay
$\tilde{\chi}^0_2 \to \tau^+\tau^- \tilde{\chi}^0_1$, with one tau decaying
leptonically, the other hadronically, leads to one identified lepton in the
final state].
Furthermore, mixing remnants with
the extra iso-singlet components eventually may make
the picture even more complicated.

However, it should be noted that neutralinos always generate the same number of
positively and negatively charged leptons.
On the other hand, chargino decay chains lead to leptons
in the final state whose charge is correlated to the squarks from which the
charginos originate.
Therefore, this scenario can
still be controlled by measuring separately the cross sections for
$\ell^+\ell^+$ and $\ell^-\ell^-$ production.
All neutralino decay modes of L- and R-squarks can be combined into one
contribution, leading to {\it equal} rates of $\ell^+$ and
$\ell^-$ in the final state signature. Then, two branching ratios remain
undetermined,
$BR(\tilde{u}_L \to \tilde{\chi}^+_1)$ and $BR(\tilde{d}_L \to
\tilde{\chi}^-_1)$, that control the {\it difference} between the rates
for $\ell^+\ell^+$ and $\ell^-\ell^-$.

This analysis relies mainly on the assumption that the chargino sector remains
unmodified with respect to the MSSM when the system is expanded. 
[Only additional 
new charginos that couple differently to $\tilde{u}_L$ and $\tilde{d}_L$ would
significantly alter that picture.] In any case, the chargino sector 
can be brought under control
from measurements at a low-energy lepton collider.
Furthermore, for our analysis of the signature of two leptons, two or more jets
and missing energy, we will obtain a decent signal only for the mass hierarchy
$m_{\go} \gg m_{\sq} \gg \mcha{1} \gg \mneu{1}$.
For different mass spectra, the selection cuts would effectively kill the SUSY
signal. It is likely that a similar analysis can
be performed in such a case by suitably adjusting the cuts, but a thorough
investigation of all possible situations is beyond the scope of this paper.

\paragraph{ } 
We repeat the analysis of the previous section, now taking as measurable
observables the [binned] $p_{\rm T}$ spectra of the 3rd and 4th jet for
$\ell^+\ell^+$ and $\ell^-\ell^-$ separately. The free parameters in the fit are
the Yukawa coupling $\hat{g}_{\rm s}$  and the
branching fractions $BR(\tilde{u}_L \to \tilde{\chi}^+_1)$ and $BR(\tilde{d}_L
\to
\tilde{\chi}^-_1)$, in addition the overall normalisation and the contribution 
from R-squark
processes, mainly $pp \to \sqL^{(*)}\sqR^{(*)}$.
[On the methodological side,
this case illustrates most clearly the technical potential of varying the
relative
weight of direct squark-pair production {\it versus} the Compton process in determining
the Yukawa coupling.] Since the closure of the minimal supersymmetric Standard Model,
in the form defined above, is a 
natural assumption, the scenario is expected to be quite general and comprehensive.
Weighing this qualification properly, the separate $\ell^+\ell^+ ,\, \ell^-\ell^-$
measurements come very close to a model-independent determination of the Yukawa coupling 
$\hat{g}_{\rm s}$.

\begin{figure}[tb]
\centering
\epsfig{figure=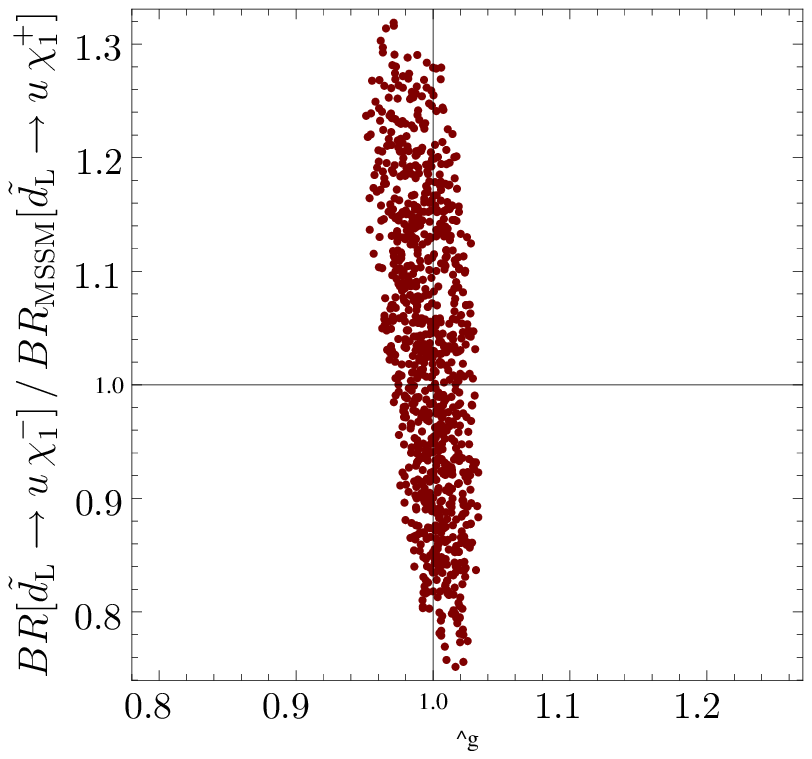, height=6cm, bb=10 445 250 678}%
\epsfig{figure=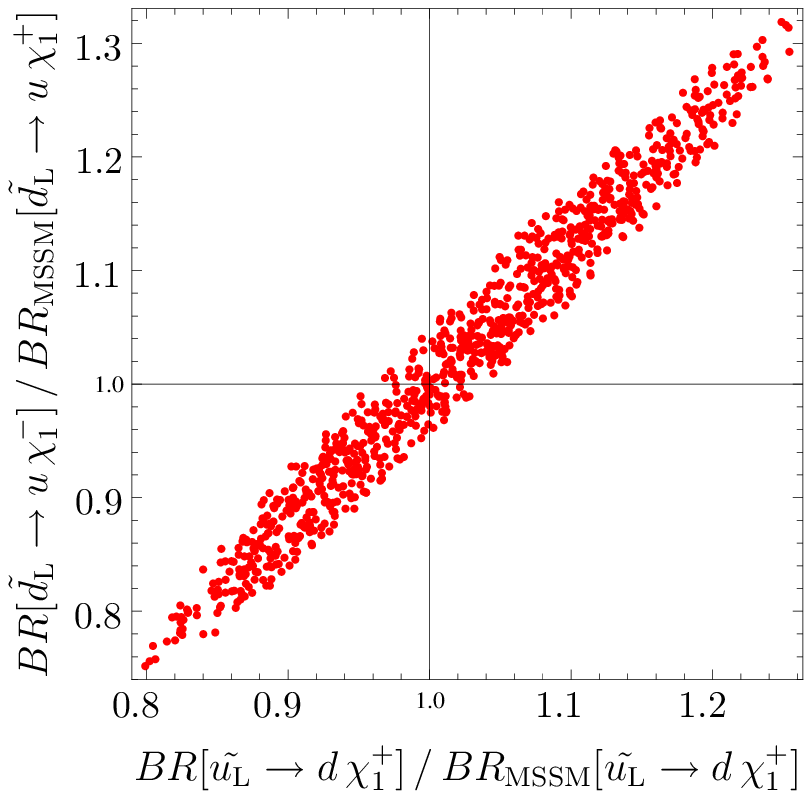, height=6cm, bb=45 445 250 678, clip=true}
\mycaption{Determination of the SUSY-QCD Yukawa coupling $\hat{g}_{\rm
s}$ and the L-squark branching ratios into charginos from a fit to the $p_{\rm
T3}$ distributions for positive ($\ell^+\ell^+$) and negative ($\ell^-\ell^-$)
lepton pairs in the final state.
The plots show the 1$\sigma$ statistical error regions corresponding to the fit
method C.}
\label{fg:plusminus}
\end{figure}

Assuming a luminosity of 300~fb$^{-1}$, a binned $\chi^2$ fit gives a statistical
error of 5.7\% for $\hat{g}_{\rm s}$, see Table~\ref{tab:err}~(Method C). The
{\it individual} squark branching ratios, on the other hand, are determined 
relatively poorly, see~Fig.~\ref{fg:plusminus}.
The origin of this uncertainty is the fact that 
both the $p_{\rm T}$ spectra as well as the ratio
of ($\ell^+\ell^+)/(\ell^-\ell^-$) signal rates discriminate {\it a priori} 
hardly between $\tilde{u}_\LL$ and $\tilde{d}_\LL$
decays. The distinction is slightly enhanced by 
the different shapes of $u$- and $d$-quark distributions in the proton,
though the fit results for $BR(\tilde{u}_L \to \tilde{\chi}^+_1)$ and
$BR(\tilde{d}_L \to \tilde{\chi}^-_1)$ remain strongly correlated.
The correlation however does not affect the accuracy of the Yukawa
coupling significantly. With a relative contribution of 12\%
to $++$ and 20\% to $--$ final states, respectively, $\tilde{\chi}^0_2$ pairs 
contribute moderately to the signal events, but they lead to almost the same
transverse jet momentum distribution as the chargino channels. This explains the
relatively precise determination of $\hat{g}_{\rm s}$, even when the squark
branching fractions are not pre-determined.

The systematic errors in the analysis method C are equivalent to Method B.
In total, the error on the ratio $\hat{g}_{\rm s}/g_{\rm s}$ 
can be expected close to 7.3\% in this largely model-independent approach. 
With a luminosity-upgraded LHC, yielding 1000
fb$^{-1}$, this error could be reduced to about 5\%.

\section{Conclusions} 

A crucial consequence of supersymmetry is the identity of gauge couplings
with the corresponding Yukawa couplings. Specifically in supersymmetric
QCD, the
quark-quark-gluon coupling and the quark-squark-gluino coupling are predicted
to be equal [up to radiative corrections for SUSY scales different
from the electroweak scale, described by renormalisation group
running]. In the present analysis we have investigated how
this relation can be tested through measurements at the
Large Hadron Collider (LHC).

After identifying a clean signal of same-sign leptons, two or more jets and
missing energy, that can be easily separated from Standard Model
backgrounds, we have elaborated how this signal receives contributions from
various squark and gluino production processes. Assuming a sufficiently large
mass difference between gluino and squarks, it was further shown how gluino and
squark production can be distinguished by analysing the transverse momentum 
distribution of additional jets.
Since the gluino and squark production cross sections depend in different ways
on the SUSY-QCD Yukawa coupling, this coupling can be extracted by measurements
of the cross sections and jet distributions. 

We have established analysis strategies that allow the extraction of the
SUSY-QCD Yukawa coupling with varying degrees of model assumptions. In the most
restricted case, it has been assumed that the complete decay pattern of the squarks
in the MSSM is known, possibly from a high-energy $e^+e^-$ collider. In this
case one can perform a measurement of the total cross sections for
squark-pair, squark-gluino and gluino-pair production at the LHC.

In a second step, we have extended the analysis to the situation when the 
squark branching ratios
are not known, allowing for a more general structure of the theory, with iso-singlet
extensions of the MSSM, for instance.
For relatively small mixing in the neutralino sector,
several useful relations between different squark partial widths can be
established based on gauge invariance, leaving
only one branching ratio, $\tilde{u}_L \to \chi^+$ to be determined 
internally within this LHC measurement.
In case of large mixing in the neutralino sector, one ends up in general with
three free parameters that nevertheless can be constrained from LHC data
such that the impact on the measurement of the SUSY-QCD Yukawa coupling 
remains under good control. 

For a specific scenario we have demonstrated the application of these methods.
We have generated Monte-Carlo samples of the squark and gluino processes as
well as the relevant backgrounds. After applying cuts to reduce the background,
we have extracted the SUSY-QCD Yukawa coupling from appropriate fits to the
simulated data. Including the most relevant expected systematic errors, we have
found that \emph{the SUSY-QCD Yukawa coupling can be determined  at LHC with a
precision of a few per-cent}, {\it i.e.} 3.4\% in the most restricted method, 
and 7.3\% in the most general method.

While these results are very encouraging, it should be pointed out that the
experimental analysis and the expected quantitative precision will depend 
on the concrete supersymmetric scenario, {\it i.e.} the sparticle mass spectrum. 
Nevertheless, our results demonstrate that precision measurements of the underlying
structure of supersymmetric theories are feasible at the LHC.
It would be interesting to refine these studies in the future by including
the impact of detector effects.

\bigskip

\vspace{- .3 cm}
\section*{Acknowledgements}

The authors are grateful to M.~M.~M\"uhlleitner for
clarifying discussions.
A.F.\ is supported in part by the Schweizer Nationalfonds. P.S.\ is
supported by 
Fermi Research Alliance, LLC, under Contract No.\
DE-AC02-07CH11359 with the United States Department of Energy and M.S.
in part by the Swiss Bundesamt f\"ur Bildung und Wissenschaft.

\end{document}